\documentclass[
reprint,
 amsmath,amssymb,
 aps,
]{revtex4-1}

\usepackage{graphicx}
\usepackage{dcolumn}
\usepackage{bm}
\usepackage[section]{placeins}
\usepackage{color}
\usepackage{mathrsfs}
\usepackage{ulem}
\usepackage{hyperref}
\usepackage{braket}

\begin{document}

\title{Optically detected magnetic resonance of nitrogen-vacancy centers in microdiamonds inside nanopolycrystalline diamond anvil cell}

\author{Masahiro Ohkuma$^{1}$}
\email{okuma.m.36c3@m.isct.ac.jp}
\author{Keigo Arai$^{1}$}
\author{Kenji Ohta$^{2}$}
\author{Toru Shinmei$^{3}$}
\author{Ryo Matsumoto$^{4}$}
\author{Yoshihiko Takano$^{4}$}
\author{Tetsuo Irifune$^{3}$}
\affiliation{$^{1}$School of Engineering, Institute of Science Tokyo, Yokohama 226-8501, Kanagawa, Japan}
\affiliation{$^{2}$Department of Earth and Planetary Sciences, Institute of Science Tokyo, Meguro 152-8551, Tokyo, Japan}
\affiliation{$^{3}$Geodynamics Research Center, Ehime University, Matsuyama, 790-8577, Ehime, Japan}
\affiliation{$^{4}$Research Center for Materials Nanoarchitectonics (MANA), National Institute for Materials Science, Tsukuba 305-0047, Ibaraki, Japan}

\date{\today}

\begin{abstract}
We demonstrated optically detected magnetic resonance (ODMR) of nitrogen-vacancy (NV) centers in microdiamonds inside a diamond anvil cell pressurized with nanopolycrystalline diamond (NPD) anvils. NPD exhibits high optical transparency, superior hardness, and low thermal conductivity, making it suitable for optical and spectroscopic measurements under high-pressure and high-temperature conditions. We observed the ODMR signal from an ensemble of NV centers under conditions where NV centers in microdiamonds served as markers for pressures exceeding 30 GPa, with a culet diameter of 600 $\mu$m. We also performed ODMR measurements on multiple microdiamonds sealed inside a sample chamber and found that the resonance frequency varied with the pressure distribution. The combination of NPD and microdiamonds containing NV centers is auspicious for pressure and magnetic sensing under concurrent high-pressure and high-temperature conditions.
\end{abstract}

\maketitle

\begin{figure}[htb!]
\centering\includegraphics[clip,width=1\columnwidth]{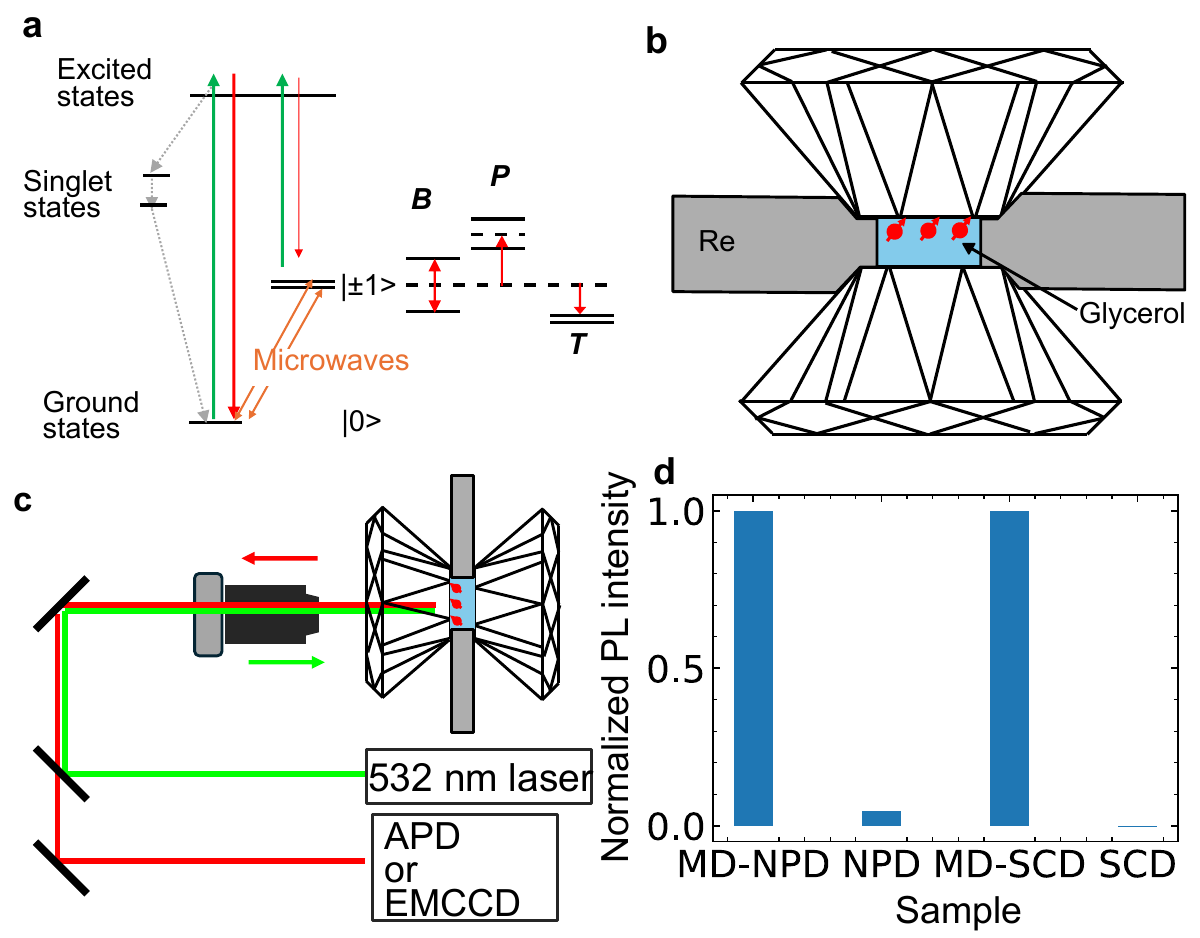}
\caption{\label{f1} (a) Energy diagram of NV center. A magnetic field $B$ along the NV center splits the $\ket{{\pm{1}}}$ states via Zeeman interaction. An isotoropic pressure $P$ and temperature $T$ shifts the ZFS in the opposite directions. Anisotropic pressure splits the $\ket{{\pm{1}}}$ states. (b) Setup of the nanopolycrystalline diamond anvil cell. Light-blue region indicates the pressure-transmitting medium (glycerin) in the sample chamber formed by a rhenium gasket. The initial gasket thickness was 250 $\mu$m and was pre-indented to 50–60 $\mu$m; the sample-chamber hole diameter was 200–300 $\mu$m. The microdiamonds are placed on the culet surface. (c) Setup for the optical measurements. DM and LP denote dichroic mirror and long pass filter, respectively. (d) Normalized photoluminescence (PL) intensity of microdiamond (MD) on nanopolycrystalline diamond (NPD) and NPD itself, compared with Type IIa single-crystalline diamond. The vertical axis is normalized to the PL intensity of the MD on the corresponding anvil: for the NPD data, to the PL intensity of the MD on NPD; and for the SCD data, to the PL intensity of the MD on SCD.}
\end{figure}

Recently, a negatively charged nitrogen-vacancy (NV) center, a point defect in diamond, has attracted considerable attention as a quantum sensor that can operate even under high pressure \cite{hsiehImagingStressMagnetism2019, lesikMagneticMeasurementsMicrometersized2019, yipMeasuringMagneticField2019, hoRecentDevelopmentsQuantum2021, bhattacharyyaImagingMeissnerEffect2024}.
The NV center carries an electronic spin of $S = 1$, whose state can be coherently manipulated using a microwave magnetic field tuned to resonance and optically read out via its red photoluminescence intensity \cite{oortOpticallyDetectedSpin1988, gruberScanningConfocalOptical1997a, jelezkoObservationCoherentOscillations2004, dohertyNitrogenvacancyColourCentre2013}. This magnetic resonance technique is known as optically detected magnetic resonance (ODMR). Zero-field magnetic resonance is observed around 2.87 GHz under ambient conditions. As shown in Fig.~\ref{f1}a, the ground state energy level is sensitive to variations in temperature, magnetic field, and pressure, enabling the nanoscale imaging of these physical quantities \cite{acostaTemperatureDependenceNitrogenVacancy2010, chenTemperatureDependentEnergy2011, toyliMeasurementControlSingle2012, kucskoNanometrescaleThermometryLiving2013, balasubramanianNanoscaleImagingMagnetometry2008, taylorHighsensitivityDiamondMagnetometer2008, lesageOpticalMagneticImaging2013, lesikMagneticMeasurementsMicrometersized2019, dohertyElectronicPropertiesMetrology2014, ivadyPressureTemperatureDependence2014, barsonNanomechanicalSensingUsing2017, hsiehImagingStressMagnetism2019, hoProbingLocalPressure2020a}. Magnetic and pressure sensing under high-pressure conditions has been extensively performed using a diamond anvil cell (DAC) as the pressure-generating apparatus. Such measurements are typically realized either by implanting NV centers near the surface of a single-crystalline diamond (SCD) anvil or by introducing nano- or micro-sized diamonds containing NV centers into the sample chamber \cite{steeleOpticallyDetectedMagnetic2017, hsiehImagingStressMagnetism2019, lesikMagneticMeasurementsMicrometersized2019, yipMeasuringMagneticField2019, hoProbingLocalPressure2020a, hoSpectroscopicStudyN$V$2023, sheltonMagnetometryDiamondAnvil2024}.
In addition, spatial pressure distributions inside DAC sample chambers have recently been investigated using NV centers in nanodiamonds \cite{sudaGPaPressureImaging2025}.

Whereas SCD is typically employed as the anvil material in DAC, nanopolycrystalline diamond (NPD) is also a viable alternative.
NPD is composed of fine grains several tens of nanometers in a randomly oriented crystallographic orientation \cite{irifuneUltrahardPolycrystallineDiamond2003}.
The optical transparency of NPD is comparable to that of typical Type Ib SCD, making it suitable for optical and spectroscopic measurements under high pressure \cite{sumiyaOpticalCharacteristicsNanoPolycrystalline2009a, fukutaElectronicPropertiesNanopolycrystalline2018}.
In addition, NPD exhibits superior hardness and isotropic mechanical properties owing to the absence of cleavage planes, enabling the generation of ultrahigh pressures, even with large culet sizes.
This capability is particularly advantageous in experiments that require large sample volumes, such as neutron diffraction \cite{sumiyaIndentationHardnessNanopolycrystalline2004, nakamotoNoteHighpressureGeneration2011, komatsuDevelopmentsNanopolycrystallineDiamond2020}.
Furthermore, NPD can exhibit a thermal conductivity nearly an order of magnitude lower than that of typical SCD, which facilitates retaining higher temperatures in the sample chamber \cite{odakePulsedLaserProcessing2009, irifune308NanopolycrystallineDiamond2014}.
When combined with laser heating, this characteristic allows the sample chamber within the DAC to be heated to temperatures exceeding 5000 K \cite{ohfujiApplicationNanopolycrystallineDiamond2010}.
In high-pressure experiments employing NPD anvils, pressure estimation based on the edge of the diamond Raman shift, commonly used with SCD anvils, is challenging \cite{irifuneFormationPurePolycrystalline2004}. Instead, pressure is typically determined using the ruby fluorescence scale.

Despite these advantages of NPD anvils, there have been no reports on their integration with NV centers, which can provide local information on the strain, magnetic field, and temperature.
However, realizing ODMR-based NV sensing in an NPD--DAC is non-trivial because NPD can introduce additional scattering and background luminescence, which can reduce fluorescence collection efficiency and ODMR contrast.
In this study, we employed microdiamonds (MDs) containing ensembles of NV centers in conjunction with NPD anvils as an alternative pressure gauge to the ruby fluorescence method.
We demonstrated the ODMR of these MDs inside a DAC pressurized using NPD anvils.
We observed the magnetic resonance of NV centers under high pressure, under conditions where NV centers in MDs served as markers for pressures exceeding 30 GPa, with a relatively large curet diameter of 600 $\mu$m.
We also performed ODMR on multiple MDs sealed inside the sample space and found that the resonance frequency varied with the pressure distribution.
These findings highlight the significant potential of combining NPD with NV-containing MDs for localized pressure and magnetic sensing under extremely high-pressure and high-temperature conditions.

High-pressure generation was achieved using the DAC.
An overview of the setup is illustrated in Fig.~\ref{f1}b.
We used two opposing NPD anvils with culet sizes of 600 $\mu$m.
A gasket of rhenium was pre-indented from 250 $\mu$m to 50--60 $\mu$m, and a 200 $\mu$m diameter hole was drilled.
Subsequently, we prepared a slit to create a gasket with a Lenz lens \cite{pravicaNuclearMagneticResonance1998a, meierMagneticFluxTailoring2017a, lesikMagneticMeasurementsMicrometersized2019}.
To prevent deformation of the slit by pressure, the sample space and slit were filled with a cBN and $\rm TiO_2$ mixture, and the gasket was pressurized.
After the pressure was released, the cBN and $\rm TiO_2$ mixture in the sample space was removed.
The sample space was then filled with a liquid pressure-transmitting medium, glycerol, and sealed with the NPD anvil, which had several MDs on the culet surface. We used MD of 15--25 $\mu$m diameter with 3.5 ppm NV center, purchased from Adamas Nanotechnologies.
Some contamination of cBN and $\rm TiO_2$ mixtures were remained in the sample space during the sealing process.
A 1.5-turn copper loop coil was placed around the anvil to apply a microwave magnetic field.
For wide-field ODMR measurements, a rhenium gasket was pre-indented from 250 $\mu$m to 50--60 $\mu$m, and a 300-$\mu$m-diameter hole was drilled.
No Lenz-lens gasket was used.
Instead, a platinum foil was placed on the culet surface to serve as a microwave antenna, without an electrical insulation layer.

Continuous-wave (CW) ODMR was performed using a custom-built optical setup. As shown in Fig.~\ref{f1}c, a 532 nm green laser (MLL-S-532B, CNI laser) was used to irradiate the MDs to excite the NV center through an objective lens (M Plan Apo SL 20x, Mitsutoyo). Red fluorescence collected through the objective lens was passed through a long-pass filter and detected by an avalanche photodiode (APD130A2, Thorlabs) coupled to a data acquisition device (NI6363, National Instruments). Microwaves were generated using a signal generator (N5171B, Keysight or SynthHD, Windfreak) and amplified using a power amplifier (ZHL-50W-63$+$, Mini-circuits). In wide-field ODMR measurements, we used EMCCD camera (ixon Ultra, Andor).
ODMR measurements were conducted at room temperature, and each ODMR spectrum was acquired within up to 10 minutes for both the APD-based and EMCCD-based measurements.

The NPD shows a broad peak in the photoluminescence (PL) spectra between 600 and 800 nm, which may be related to the presence of defects, dislocations, and grain boundaries \cite{sumiyaOpticalCharacteristicsNanoPolycrystalline2009a, fukutaElectronicPropertiesNanopolycrystalline2018}.
Therefore, the signal detected using the avalanche photodiode includes a background signal derived from the NPD.
Figure~\ref{f1}d shows the PL intensities of the anvil with an MD and the anvil alone for both NPD and Type IIa SCD, where each dataset is normalized to the PL intensity of the corresponding anvil with MD configuration.
The PL intensity from NPD was 4$\%$ of the MD, such that NPD did not affect the ODMR experiments in our setup.
On the other hand, background PL from NPD may become a limiting factor for future applications that employ weaker emitters (e.g., nanodiamonds and/or lower NV concentrations), where the signal-to-background ratio can decrease.
For comparison, the background signal from the Type IIa SCD alone was below the APD detection limit.

\begin{figure}[htb!]
\centering\includegraphics[clip,width=1\columnwidth]{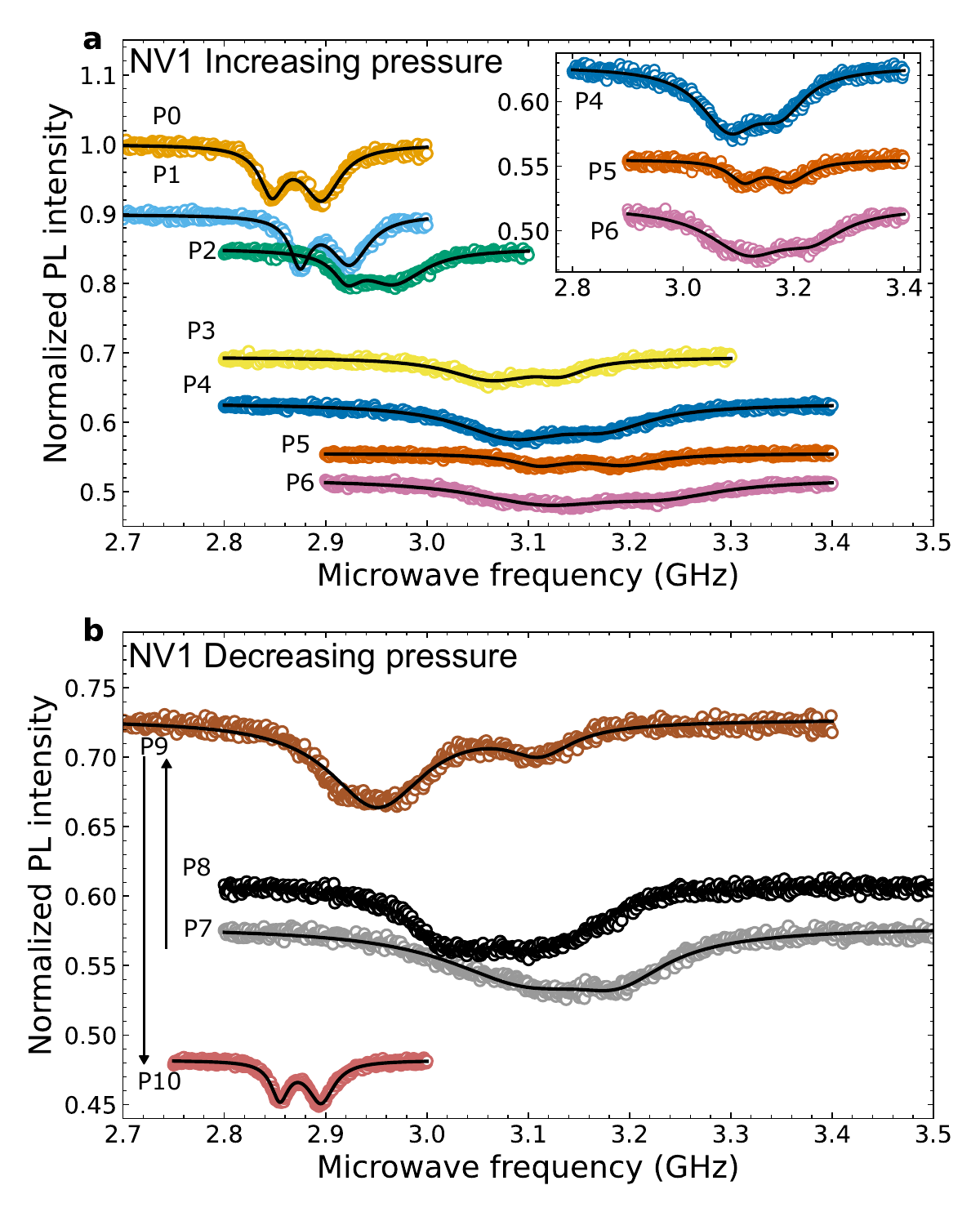}
\caption{\label{f2} CW ODMR spectra of the MD labeled NV1 under high pressure with (a) increasing pressure process and (b) decreasing pressure process. The vertical axes are normalized by the PL intensity of the ambient pressure, labeled as P0. The black lines indicate fitting curves.
The black arrows in (b) indicate the measurement order.}
\end{figure}

CW ODMR was performed at increasing pressure (P0--P6) and decreasing pressure (P7--P10).
Figure~\ref{f2} shows the CW ODMR spectra of the MD labeled NV1 at a laser power of 5 mW.
The vertical axis was normalized to the PL intensity at ambient pressure, labelled P0.
At ambient pressure, dips corresponding to the magnetic resonance were observed.
The resonance peak split into two, which can arise from local strain and/or local electric fields associated with charged defects or impurities in the MDs \cite{mittigaImagingLocalCharge2018, yuOpticallyDetectedMagnetic2024}.
By increasing the pressure, the resonance peaks shifted toward higher frequencies, and the PL intensity decreased, as previously reported \cite{dohertyElectronicPropertiesMetrology2014, daiOpticallyDetectedMagnetic2022}.
The ODMR spectra were recorded during the depressurization process to ambient pressure.
The PL intensity recovered with decreasing pressure, whereas the PL intensity at P10 (after full pressure release) was approximately $50\%$ of that before applying the pressure. It is considered that the MDs moved when the pressure was released, causing some of the fluorescence to be blocked by the cBN and $\rm TiO_2$ mixture.

In P9, the difference between the two resonance peaks became larger than those in P7 and P8.
This provides a plausible scenario in which P7/P8 reflect a broader distribution of local strain environments within the MD, leading to spectral broadening.
By contrast, in P9, stress/contact redistribution during decompression may render narrower local strain environments more dominant in the detected signal, allowing the splitting to be resolved more clearly.
The ODMR lineshape in Fig.~\ref{f2} varies with pressure, which may reflect changes in relaxation processes as well as inhomogeneous broadening due to local stress/strain.
A more quantitative decomposition of these contributions using established zero-field ODMR lineshape analysis methods is an important next step \cite{hayashiOptimizationTemperatureSensitivity2018}.

\begin{figure}[htb!]
\centering\includegraphics[clip,width=1\columnwidth]{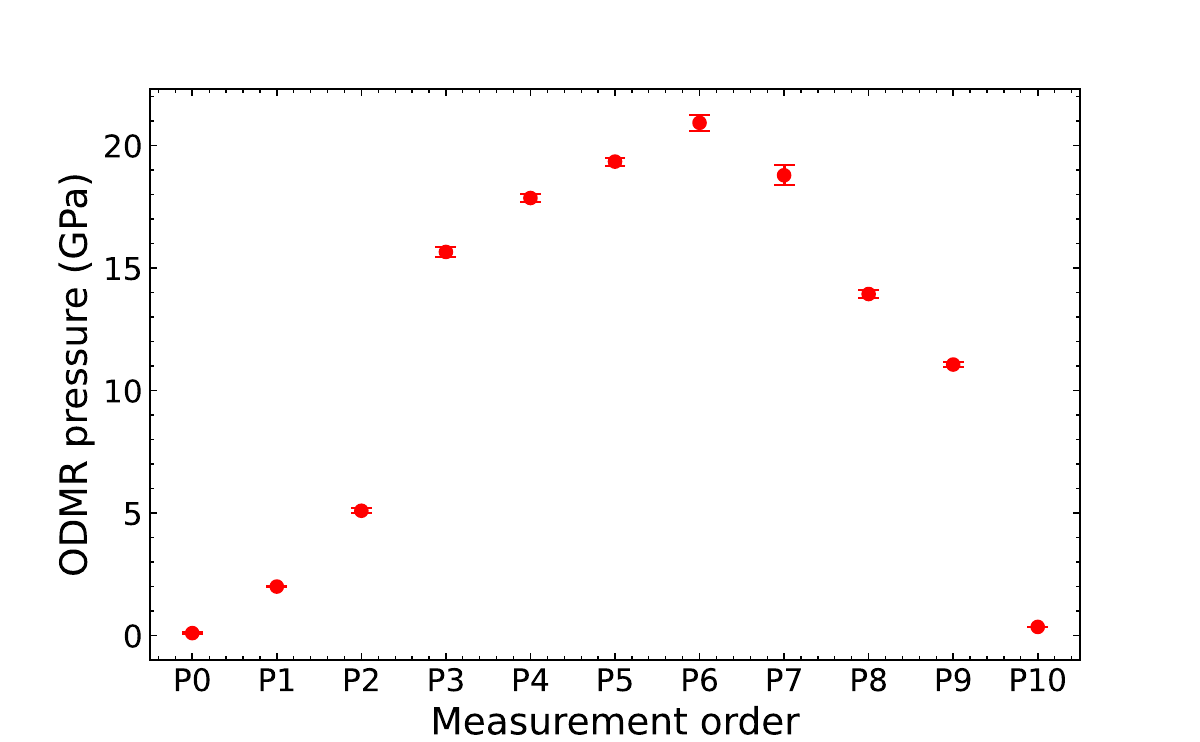}
\caption{\label{f3} Pressure estimated from ODMR at each measurement sequence. Pressures are estimated using $P_{\rm ODMR}=(D-D_0)/\beta$ GPa, where $D_0=2.87\times10^9$ Hz, $\beta = 14.58\times10^6$ GPa/Hz, and $D$ denotes resonance frequency \cite{dohertyElectronicPropertiesMetrology2014}. The error bar represents a 95$\%$ confidence interval of the fitted value of $D$.}
\end{figure}

The pressure was estimated from the value of the resonance frequency $D$ under high pressure, using $P_{\rm ODMR}=(D-D_0)/\beta$ GPa, where $D_0=2.87$ GHz and $\beta = 14.58$ MHz/GPa \cite{dohertyElectronicPropertiesMetrology2014}.
Figure~\ref{f3} shows the ODMR pressures for each measurement.
We successfully performed ODMR above 20 GPa. Because the ODMR contrast is approximately 8$\%$ even at 20 GPa, it is possible to perform ODMR even at high temperatures where the PL intensity and contrast decrease \cite{toyliMeasurementControlSingle2012}.

\begin{figure}[htb!]
\centering\includegraphics[clip,width=1\columnwidth]{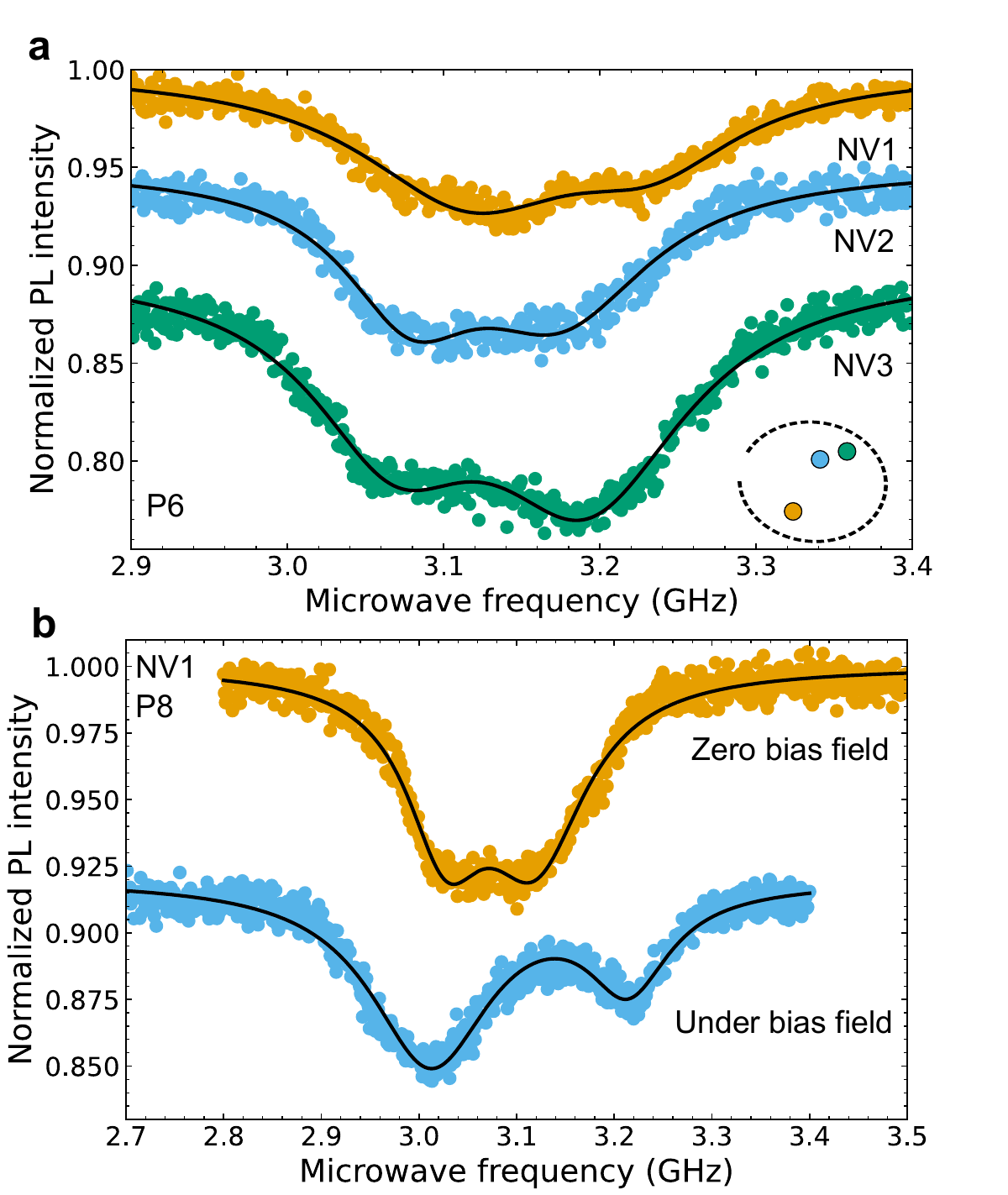}
\caption{\label{f4} (a) CW ODMR spectra on three MDs with NV center, termed as NV1, NV2, and NV3, respectively. The results of P6 are presented in this section. NV1 is the same as that used in Figs~\ref{f2} and \ref{f3}. These spectra have offsets of $-0.05$ for NV2 and $-0.1$ for NV3, respectively. The inset shows the position of each MD in the sample space.
(b) CW ODMR spectra on NV1 at pressure sequence of P8 with no bias field and finite bias field. The spectrum obtained under finite bias field has offsets of $-0.08$.}
\end{figure}

In high-pressure experiments using a DAC, a pressure distribution is expected, especially when the pressure medium is a solid or becomes solid under pressure.
Because MDs can be dispersed throughout the sample space, both the local pressure and the magnetic field can be detected, providing beneficial insights into materials under high pressure \cite{yipMeasuringMagneticField2019, hoProbingLocalPressure2020a, hoSpectroscopicStudyN$V$2023}.
Here, we demonstrate the ODMR of three MDs, NV1, NV2, and NV3. NV1 is the same as that used in Figs~\ref{f2} and \ref{f3}. Figure~\ref{f4} shows the ODMR results for the MDs and their positions.
The resonance frequencies and pressure estimated from ODMR were $3.175\pm0.005$  GHz and $20.9\pm0.3$ GPa, $3.127\pm0.004$ GHz and $17.6\pm 0.3$ GPa, and $3.130\pm0.002$ GHz and $17.8\pm0.1$ GPa for NV1, NV2, and NV3, respectively.

In addition to pressure manometry, the present platform may provide a route toward magnetic sensing under high pressure.
To demonstrate feasibility under a bias field, we acquired ODMR spectra under an externally applied magnetic field.
Figure~\ref{f4}(b) shows the ODMR spectra of NV1 at P8 under the applied magnetic field.
The bias magnetic field was applied by a permanent magnet.
The ODMR spectra showed clear field-dependent spectral changes even in the NPD-anvil configuration.
This result supports the applicability of the platform to magnetic measurements in high-pressure environments, as also suggested by prior demonstrations of magnetic sensing using NV centers hosted in diamond particles under pressure \cite{yipMeasuringMagneticField2019}.

\begin{figure}[htb!]
\centering\includegraphics[clip,width=1\columnwidth]{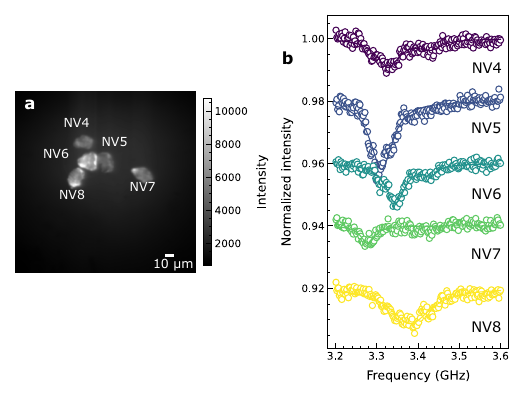}
\caption{\label{f5} (a) Fluorescence image of the MDs captured by the EMCCD camera. (b) CW ODMR spectra under high pressure. The spectra are offset by $0.02$ each for clarity.}
\end{figure}

We also performed wide-field ODMR measurements, which is a standard technique for simultaneously observing the entire sample chamber in a single shot.
Five MDs, labeled NV4 to NV8, were enclosed within the sample chamber.
Figure \ref{f5}(a) shows a fluorescence image of the MDs captured by the EMCCD camera.
Figure \ref{f5}(b) presents the ODMR spectra obtained under high pressure.
The plotted ODMR data represents the signal intensity of a pixel for each respective microdiamond.
The resonance frequencies of the microdiamonds ranged from 3.28 GHz to 3.39 GHz.
These values correspond to pressures between 28 GPa and 35 GPa.

Pressures inferred from different MDs can differ by more than 10$\%$.
Such a spread is naturally expected under quasi-hydrostatic conditions once the pressure-transmitting medium solidifies, as spatial pressure inhomogeneity and deviatoric stress develop across the sample chamber.
Indeed, Klotz \textit{et al.} quantified pressure gradients by distributing multiple ruby spheres and analyzing the standard deviation of ruby-derived pressures, reporting, for example, a standard deviation of $\sim$2~GPa at $\sim$20~GPa for a methanol--ethanol--water mixture \cite{klotzHydrostaticLimits112009}.
Accordingly, pressures obtained from individual MDs should be interpreted as local effective pressures rather than a single uniform chamber pressure.
In future studies, simultaneous ruby fluorescence measurements would provide a useful independent cross-check of the pressure scale and its spatial distribution.

The formation of the NV center and manipulation of its spin state in NPD remain challenging.
Although a PL peak near 637 nm has been observed in the as-grown NPD at 5 K, suggesting the natural presence of negatively charged NV centers, the formation and stability of these centers at room temperature remain uncertain \cite{fukutaElectronicPropertiesNanopolycrystalline2018}.
NPD contains approximately 100 ppm of nitrogen, whereas the P1 center concentration is below 1 ppm \cite{sumiyaOpticalCharacteristicsNanoPolycrystalline2009a}. It is of interest to investigate whether NV centers can be introduced and stabilized in NPD via ion implantation or chemical vapor deposition methods, as demonstrated in SCDs \cite{barrySensitivityOptimizationNVdiamond2020}.

A key practical advantage of NPD anvils is that they can generate high pressures while maintaining relatively large culet sizes, which generally translates into a larger available sample chamber volume \cite{nakamotoNoteHighpressureGeneration2011}.
In the present experiments, we used a 0.6~mm culet, and previous studies have reported that millimeter-class culets can still reach pressures up to $\sim$80~GPa \cite{komatsuDevelopmentsNanopolycrystallineDiamond2020}.
A larger accessible volume is advantageous for (i) the simultaneous measurement of multiple samples within a single loading and (ii) future integration with probes that benefit from larger sample volumes, such as neutron scattering.
Importantly, achieving ODMR in an NPD--DAC is not guaranteed a priori because NPD can introduce additional scattering and background luminescence that may degrade the ODMR contrast.
Therefore, our results provide a practical foundation for extending NV-based sensing to extreme-pressure experiments, where larger sample volumes and multi-sample throughput are desirable.

The manipulation of the NV center under the concurrent application of high pressure and high temperature presents a compelling area of study because the NV center is known to function at high temperatures of 1400 K \cite{liuCoherentQuantumControl2019, fanQuantumCoherenceControl2024} or high pressures of up to 100 GPa \cite{daiOpticallyDetectedMagnetic2022, bhattacharyyaImagingMeissnerEffect2024, wangImagingMagneticTransition2024}. Owing to the exceptionally high thermal conductivity of SCD, maintaining a higher temperature within the sample chamber presents significant challenges. In contrast, NPD exhibits lower thermal conductivity, thereby facilitating the retention of higher temperatures \cite{odakePulsedLaserProcessing2009}.
Consequently, NPD is advantageous for conducting magnetic research under extreme conditions of high pressure and high temperature.
On the other hand, ODMR contrast decreases substantially at elevated temperatures and under non-hydrostatic stress, and the resulting resonance broadening/splitting can hinder quantitative analysis.
Accordingly, performing ODMR-based NV sensing center under such conditions remains challenging.
Employing hydrostatic loading with a fluid pressure-transmitting medium would help mitigate deviatoric stress and improve the reliability of magnetic sensing.

While the present results support the feasibility of ODMR-based sensing in the NPD-DAC geometry, extending the platform to quantitative magnetic-field imaging requires additional considerations when using MDs.
Magnetic-field imaging using randomly oriented micro-/nano-diamond sensors has been reported under ambient conditions \cite{sengottuvelWidefieldMagnetometryUsing2022,tsukamotoAccurateMagneticField2022}, indicating that imaging is feasible in principle even without a predefined crystal orientation.
In our experiments, individual MDs are generally placed with orientations that are random with respect to the applied bias field.
Under a bias field, this leads to MD-dependent ODMR spectral patterns because multiple NV orientations contribute distinct resonances that can partially overlap, reducing practical resolvability and complicating quantitative fitting.
Furthermore, under DAC conditions, non-hydrostatic stress can induce additional resonance broadening and/or splitting of the resonances and reduce the ODMR contrast, making quantitative analysis even more challenging.
Therefore, more quantitative magnetic imaging in future work would benefit from hydrostatic pressure conditions using a gas pressure-transmitting medium and/or sensor geometries with a well-defined orientation, for example, by employing an SCD sensor chip inserted into the sample chamber, as adopted in previous studies \cite{wangAcSensingUsing2021}.

Recently, some authors have developed an electrical transport measurement system under high pressure, using boron-doped diamond (BDD) as electrodes \cite{matsumotoNoteNovelDiamond2016, matsumotoDiamondAnvilCells2018, matsumotoElectricalTransportMeasurements2020}. Additionally, they have engineered a heating system for high-pressure environments employing BDD as a resistance heater and a thermometer \cite{matsumotoDiamondAnvilCell2021, matsumotoHighPressureSynthesisSuperconducting2022, matsumotoEmergenceSuperconductivity202025}. BDD can be synthesized on NPD through microwave plasma chemical vapor deposition, a process similar to that used for SCD. The integration of BDD electrodes, BDD heaters, and MDs with NV centers facilitates the simultaneous measurement of electrical resistance and magnetic fields under conditions of elevated temperature and pressure. Very recently, we have demonstrated coherent control of NV spins using a BDD circuit as a microwave antenna, which could work under high pressure and temperature \cite{ohkumaCoherentControlSolidstate2024}. The BDD electrode is highly durable and can be reused multiple times until the diamond itself breaks, reducing experimental preparation time and subsequently facilitating the investigation of physical properties under high-temperature and high-pressure conditions.

In conclusion, we demonstrated continuous-wave optically detected magnetic resonance of nitrogen vacancy centers in microdiamonds inside a diamond anvil cell pressurized with nanopolycrystalline diamonds as the anvil. We observed magnetic resonance on nitrogen vacancy centers under conditions where the nitrogen vacancy centers in microdiamonds served as markers for pressures exceeding 30 GPa.
Furthermore, we observed continuous-wave optically detected magnetic resonance in three independent microdiamonds within the sample space, where the resonance frequency of the NV center varied with the pressure distribution. The combination of nanopolycrystalline diamonds and microdiamonds containing NV centers is up-and-coming for pressure and magnetic sensing under concurrent high pressure and temperature.

\begin{acknowledgments}
 This work was supported by JSPS KAKENHI Grant numbers JP24KJ1035, JP23K26528, JP23KK0267. This work was also supported by the Joint Usage/Research Center PRIUS, Ehime University, Japan. M.O. receives funding from JSPS Grant-in-Aid for JSPS Fellows Grant number JP24KJ1035.
\end{acknowledgments}
\bibliography{NPD}

\end{document}